\documentclass[11pt]{article}

\usepackage[margin=1in]{geometry}
\usepackage{graphicx}

\def\beq{\begin{equation}}
\def\eeq#1{\label{#1}\end{equation}}
\def\eeqn{\end{equation}}

\def\beqa{\begin{eqnarray}}
\def\eeqa#1{\label{#1}\end{eqnarray}}
\def\eeqan{\end{eqnarray}}

\let\bar=\overbar

\def\Dslash{\not{\hbox{\kern-4pt $D$}}}
\def\dslash{\not{\hbox{\kern-2pt $\del$}}}

\def\msb{{\bar{\ssstyle M \kern -1pt S}}}

\def\Title#1{\begin{center} {\Large {\bf #1} } \end{center}}
\def\Author#1{\begin{center} {\normalsize {\sc #1} } \end{center}}
\def\Institution#1{\begin{center} {\normalsize {\it #1} } \end{center}}
\def\Abstract#1{\noindent {\normalsize {\bf Abstract:} {\normalfont #1}}}
\def\Conference{\vspace{4mm}\begin{raggedright} {\normalsize {\it Talk presented at the 2019 Meeting of the Division of Particles and Fields of the American Physical Society (DPF2019), July 29--August 2, 2019, Northeastern University, Boston, C1907293.} } \end{raggedright}\vspace{4mm}}

\begin{document}

%
%

\Title{The Innovative Design of the Endcap Disc DIRC Detector for PANDA at FAIR}

\Author{M.~Schmidt, M.~D\"uren, E.~Etzelm\"uller, K.~F\"ohl, A.~Hayrapetyan, I.~K\"oseoglu, K.~Kreutzfeld, J.~Rieke}

\Institution{II. Physikalisches Institut, Justus Liebig University of Giessen, Giessen, Germany}

\Author{A.~Ali, A.~Belias, R.~Dzhygadlo, A.~Gerhardt, M.~Krebs, D.~Lehmann, K.~Peters, G.~Schepers, C.~Schwarz, J.~Schwiening, M.~Traxler}

\Institution{GSI Helmholtzzentrum f\"ur Schwerionenforschung GmbH, Darmstadt, Germany}

\Author{L.~Schmitt}

\Institution{FAIR, Facility for Antiproton and Ion Research in Europe, Darmstadt, Germany}

\Author{M.~B\"ohm, A.~Lehmann, M.~Pfaffinger, S.~Stelter, F.~Uhlig}

\Institution{Friedrich Alexander University of Erlangen-Nuremberg, Erlangen, Germany}

\Author{C.~Sfienti}

\Institution{Institut f\"{u}r Kernphysik, Johannes Gutenberg University of Mainz, Mainz, Germany}

\Abstract{The key component of the future PANDA experiment at FAIR is a fixed-target detector for collisions of antiprotons with a proton target up to a beam momentum of 15\,GeV/c and is designed to address a large number of open questions in the hadron physics sector. In order to guarantee an excellent PID for charged hadrons in the polar angle range between $5^\circ$ and $22^\circ$, a new type of Cherenkov detector called Endcap Disc DIRC (EDD) has been developed for the forward endcap of the PANDA target spectrometer. The desired separation power of at least 3\,s.d. for the separation of $\pi^\pm$ and $K^\pm$ up to particle momenta of 4\,GeV/c was determined with simulation studies and validated during various testbeam campaigns at CERN and DESY.}

\Conference

%
%

\section{Introduction}
The future PANDA detector at the Facility of Antirproton and Ion Research (FAIR) in Darmstadt, Germany, is designed as a fixed target detector for collisions of an antiproton beam with a proton target \cite{DESTEFANIS2013199}. After their generation by collisions of protons with a Nickel target, these antiprotons will be stored in the High Energy Storage Ring (HESR) which uses a pickip and kicker for the stochastical cooling of the antiproton beam before the collision in PANDA. The HESR can run in two different modes: A high resolution mode, that will be chosen in the starting phase of PANDA, and a high luminosity mode with a maximum luminosity of
\begin{equation}
\mathcal{L} = 2\cdot 10^{32}\,\mathrm{cm}^{-2}\mathrm{s}^{-1}
\end{equation}
This luminosity corresponds to an interaction rate of $\dot{N}=20$\,MHz and can be achieved by constructing another storage ring called Recycling Energy Storage Ring (RESR) after the 2nd phase of PANDA.

The PANDA detector is going to consist of two spectrometers: one target spectrometer, designed as an onion shell like detector around the interaction point, and one forward spectrometer that is going to  cover the small polar angles up to $5^\circ$ in vertical and $10^\circ$ in horizontal direction.
\begin{figure}
\begin{center}
\includegraphics[width=0.9\textwidth]{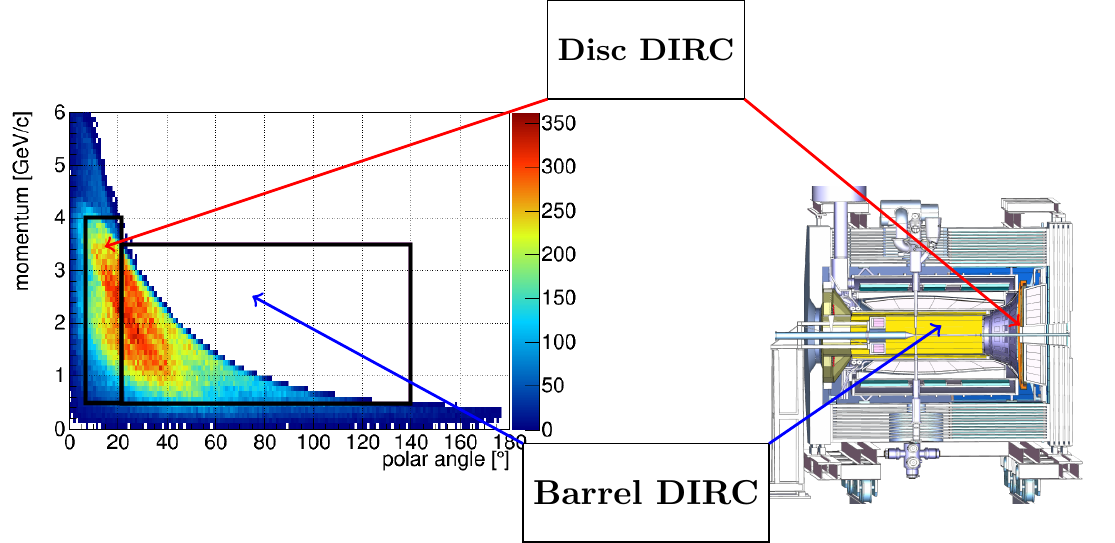}
\end{center}
\caption{\label{fig:spectrometer}The PANDA target spectrometer with the highlighted Endcap Disc DIRC (orange) and Barrel DIRC (yellow) together with their coverage of the simulated kaon phase space in PANDA.}
\end{figure}

In order to measure special resonances precisely, the particle beam momentum of PANDA can be varied between 1.5 and 15\,GeV/c. In general, all physics programs at PANDA require an excellent particle identification (PID) for charged pions and kaons. For that purpose, two Cherenkov detectors have been developed for the PANDA target spectrometer. Both detectors are based on the detection of internally reflected Cherenkov light (DIRC). The Endcap Disc DIRC (EDD) will be placed at the forward endcap of the PANDA target spectrometer with a distance of 2\,m from the proton target and is going separate pions and kaons with momenta between 1.5\,GeV/c and 4\,GeV/c inside the polar angle range from $5^\circ$ to $22^\circ$ \cite{Schmidt:2018fbc}. Larger polar angles between $22^\circ$ and $140^\circ$ are going to be covered by the Barrel DIRC that is designed in a barrel shape around the interaction point. It will cover a similar momentum range  from 1.5\,GeV/c to 3.5\,GeV/c \cite{Schwarz:2019law}. A detailed sketch of the PANDA target spectrometer with the kaon phase space covered by the two DIRC detectors is shown in Figure~\ref{fig:spectrometer}.

\section{The Endcap Disc DIRC}
\begin{figure}
\begin{center}
\includegraphics[width=0.9\textwidth]{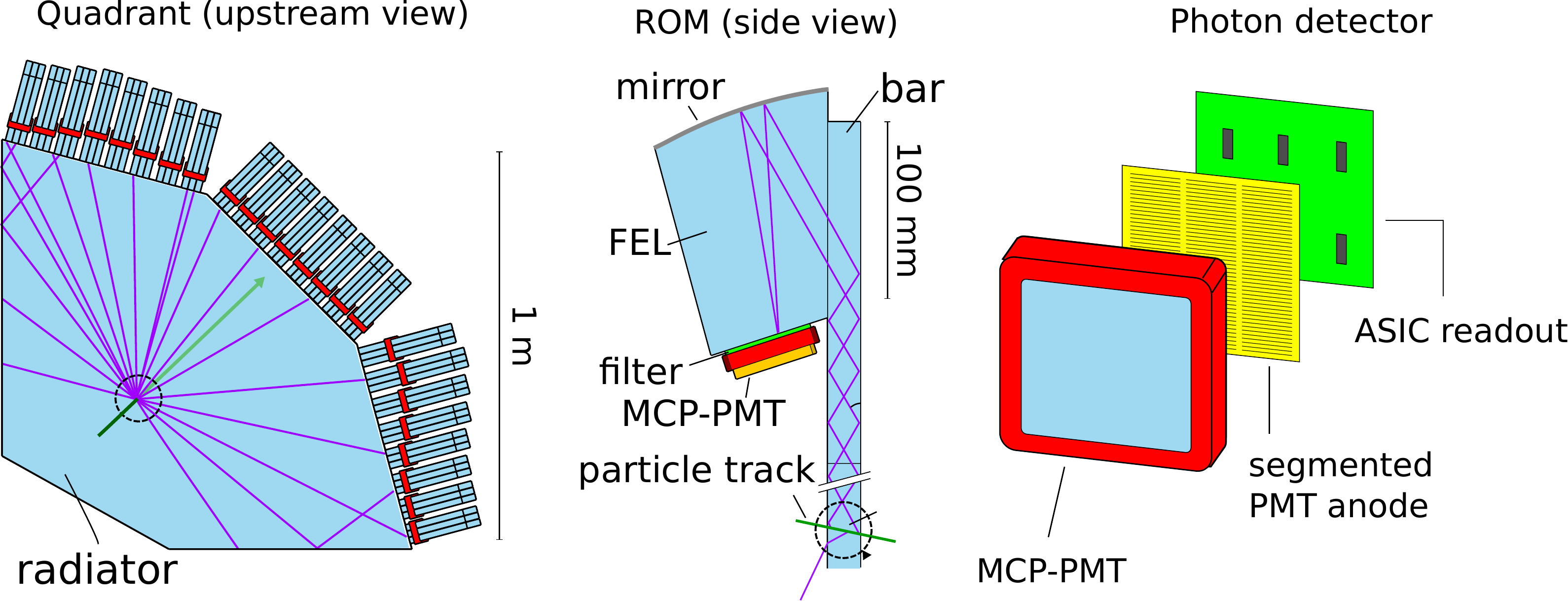}
\end{center}
\caption{\label{fig:disc}Left: A sketch of one radiator quadrant indicating a traversing charged particle (green) and the created Cherenkov photons (purple) that are propagating to the FELs at the rim. Center: The shape of one FEL including the bar and the cylindrical mirror. In addition to that, the position of the filter and the attached MCP-PMT is shown. Right: The MCP-PMT photo cathode together with the segmented anode (300 pixel) and a possible readout PCB.}
\end{figure}
The EDD, as displayed in the sketch in Figure~\ref{fig:disc}, consists of a thin fused silica radiator plate with a thickness of 2\,cm that is subdivided into 4 independent quadrants \cite{phd:etzelmueller}. If a charged particle traverses the radiator with a speed that is faster than the phase speed of light in fused silica, Cherenkov light is emitted \cite{paper:cherenkov}.
The value of the Cherenkov angle depends on the particle speed and the refractive index of the radiator material. In the EDD, the Cherenkov photons propagate via internal reflections to the outer rim of the radiator quadrant where they enter one of the 288 focusing elements (FEL). Each FEL consists of a fused silica bar, which is glued to the radiator quadrant, and a cylindrical mirror at its backside. The purpose of the mirror is to focus the Cherenkov photons on the photocathode of an attached Multichannel Plate Photomultiplier Tube (MCP-PMT) which is used to measure the time and the position information of each photon hit individually. 3 FELs each are grouped together to a readout module (ROM) and share one MCP-PMT. An additional optical filter can be used to reduce the dispersion effects of the radiator material that usually deteriorate the Cherenkov angle resolution. The necessity and wavelength interval of this bandpass filter depends mainly on the quantum efficiency (QE) curve of the chosen MCP-PMT.

In combination with the trajectory of the charged particle, the acquired hit pattern is used to reconstruct the Cherenkov angle and calculate the likelihood value for each particle species with a sophisticated reconstruction algorithm. The performance goal of the EDD is to separate charged pions and kaons with a separation power of at least 3 standard deviations (s.d.) over the full phase space. This requires an overall detector resolution of at least 2.1\,mrad which results from theoretical calculations of the Cherenkov angle difference between these two particle species.

\begin{figure}
\begin{center}
\null\hfill
\includegraphics[width=0.35\textwidth]{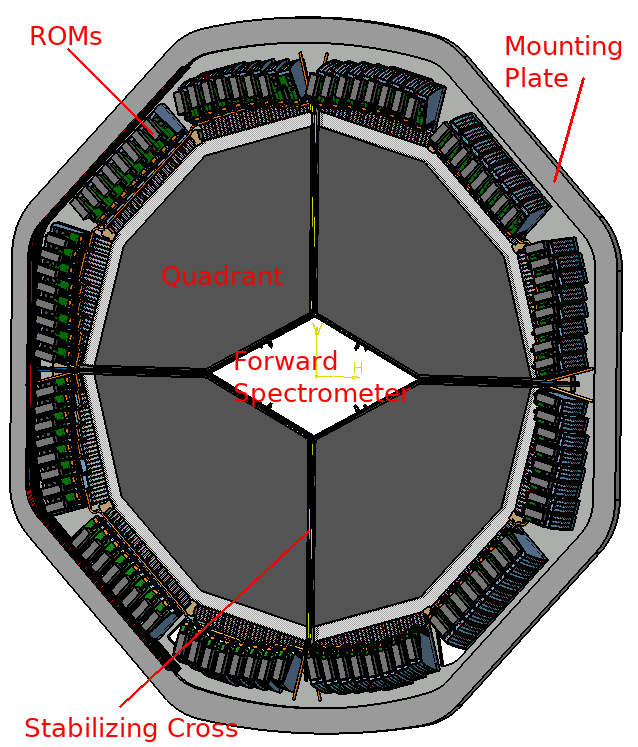}\hfill
\includegraphics[width=0.35\textwidth]{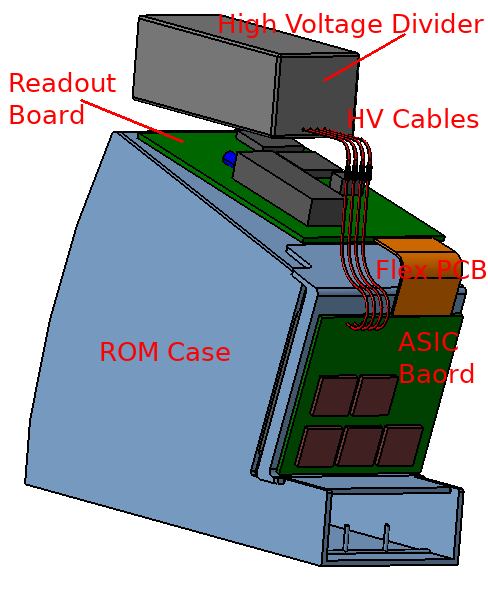}
\hfill\null
\end{center}
\caption{\label{fig:specifications}Left: A CAD drawing of the full EDD including mounting plate and stabilizing cross. Right: A CAD drawing of a ROM case in combination with the PCB, voltage divider and low/high voltage cables.}
\end{figure}

The four quadrants will be attached to a mounting plate, as shown on the left side of Figure~\ref{fig:specifications}. It will be fixed to the neighboring endcap electromagnetic calorimeter (EMC) that is going to be installed behind the EDD in upstream direction. An additionally attached stabilizing cross is used as a support structure for the quadrants. The positions of the ROMs are uniformly distributed around the radiator rim, i.e. 8 ROMs are attached to every side. The width of the FEL has been defined to 16\,mm in order to fulfill the detector resolution requirements for large polar angles. In addition to that, a gap of 1\,mm between each 2 FELs have to be taken into account. The 3 FELs of one ROM will be fixed inside a dedicated ROM holder. A sketch of this holder is presented on the right side of Figure~\ref{fig:specifications}. The electronics of the readout system is going to be attached on the ROM case. Additionally, this case has to be sufficiently light-tight in order to reduce to the amount of photon background.

The readout system consists of TOFPET ASICs \cite{article:tofpet} from the company PETsys with an intrinsic time resolution of approx. 30\,ps to digitize the sensor signal. One ASIC is used to read out 64 channels (MCP-PMT channels) simultaneously. Hence, 5 ASICs are required to read out all 300 pixels of each MCP-PMT. This results in 28,800 connected MCP-PMT channels for the data taking process in the final detector. The digital signals are delivered via traces in a flex PCB to a readout board that contains an FPGA and a bidirectional optical connection for sending the collected data to a data concentrator. Additionally, this connection will be used to provide the time synchronization by the SODAnet protocol \cite{Kuhn:2017bbd}. The high voltage for the MCP-PMTs is going to be provided by a dedicated resistor based voltage divider, whereas the low voltages for the readout electronics will be provided by external power supplies and DC-DC converters on the readout board. Furthermore, a cooling pipe for for transportation of the heat dissipation of the ASICs and FPGA chip and an optical fiber for laser measurements are foreseen in the CAD drawings of the final design.

Detailed Monte-Carlo studies have shown that the creation of around 1 charged track per collision per quadrant is expected at the highest antiproton beam momentum. It was calculated and proven with Geant4 \cite{article:geant4} simulations that each particle creates approx. 1000 Cherenkov photons for $\beta\approx 1$. Theoretical calculations have pointed out that 70\% to 80\% photons fulfill the condition of internal reflection and are thus trapped inside the radiator plate. Taking all absorption losses and efficiencies into account, in average 22 photon hits per track are going to remain. Hence, assuming an interaction rate of 20\,MHz in the high luminosity mode of PANDA leads to a hit frequency of 60\,kHz per channel. This data has to be transferred via a data concentrator to the PANDA data acquisition (DAQ) system which is currently under development. The maximum hit occupancy of one channel had been measured in Giessen with laser pulses to around 100\,kHz. It is therefore expected that even in the high luminosity mode of PANDA the estimated hit rate should not exceed the given specifications.
\section{Simulation Studies}
Simulation studies in a dedicated PANDA framework called PandaRoot \cite{article:spataro_pandaroot} have been performed in order to evaluate the possible separation between pions and kaons. For that purpose, a large amount random probe tracks with uniformly distributed azimuth and polar angles were created. Additionally, the $\cos\theta$ effect had been taken into account by enlarging the number of tracks depending on the simulated polar angle. The particle propagation is simulated with the help of the implemented Geant4 MC framework inside PandaRoot including all important optical and mechanical parameters. The MC truth information were stored and smeared according to the estimated tracking resolution of the PANDA GEM detector in front of the endcap in order to use these information directly in the reconstruction procedure.

The binning of the track position on the radiator disk surface was done in the reconstruction stage. A very fine binning of around 2\,mm has been chosen in order to study the uniformity of the misidentification in detail. The results show, as previously expected, a significantly higher separation power for lower particle momenta. This is a direct consequence of the huge differences in the rest mass of both particle species. For momenta around 4\,GeV/c, the separation power reaches values up 7\,s.d. in the regime of small polar angles. At the polar angle boundary near $22^\circ$, the separation power drops to almost 3\,s.d. This behavior can be explained with the increasing geometrical error for particles that enter the radiator close to the FELs. The results for polar angles larger than $22^\circ$ have not been considered, since this area is shielded by the barrel EMC in the PANDA target spectrometer and will not be used in final experiment either. To sum up, the obtained values for all polar angle and momentum combinations fulfill the required PANDA conditions.

In addition to the expected behavior, the presented results indicate an area of inefficiency close to the lower edge of the radiator. This drop in the separation power is a result of an overlap of different hits related to the direct and reflected hit pattern. The time resolution of the readout system is not high enough to provide a sufficient separation of particle species in this area. Because the helix radius of the charged track gets larger for a smaller magnetic field, this inefficient area shifts upwards to larger $y$ values at smaller momenta.
\begin{figure}
\begin{center}
\includegraphics[width=0.49\textwidth]{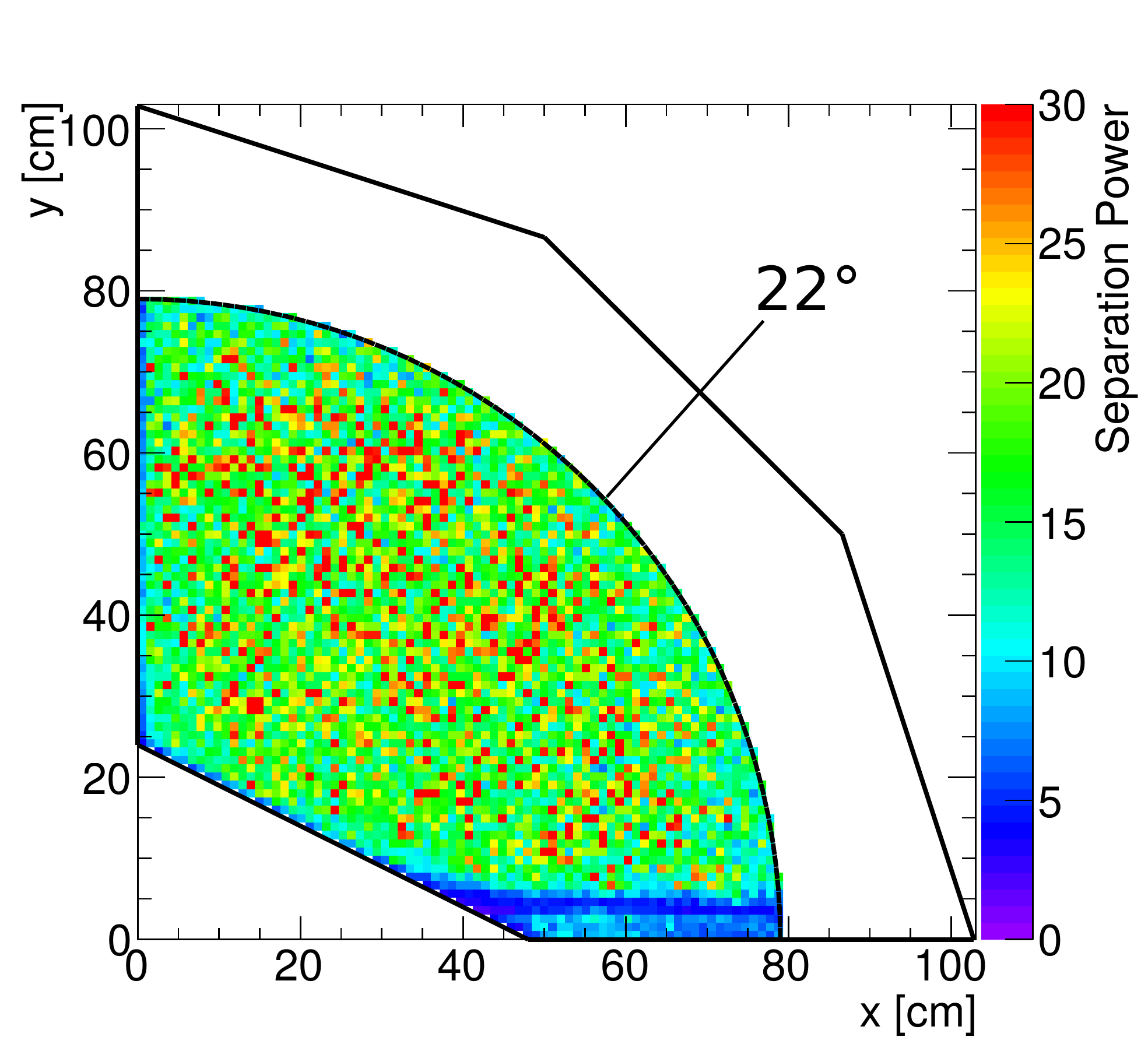}
\includegraphics[width=0.49\textwidth]{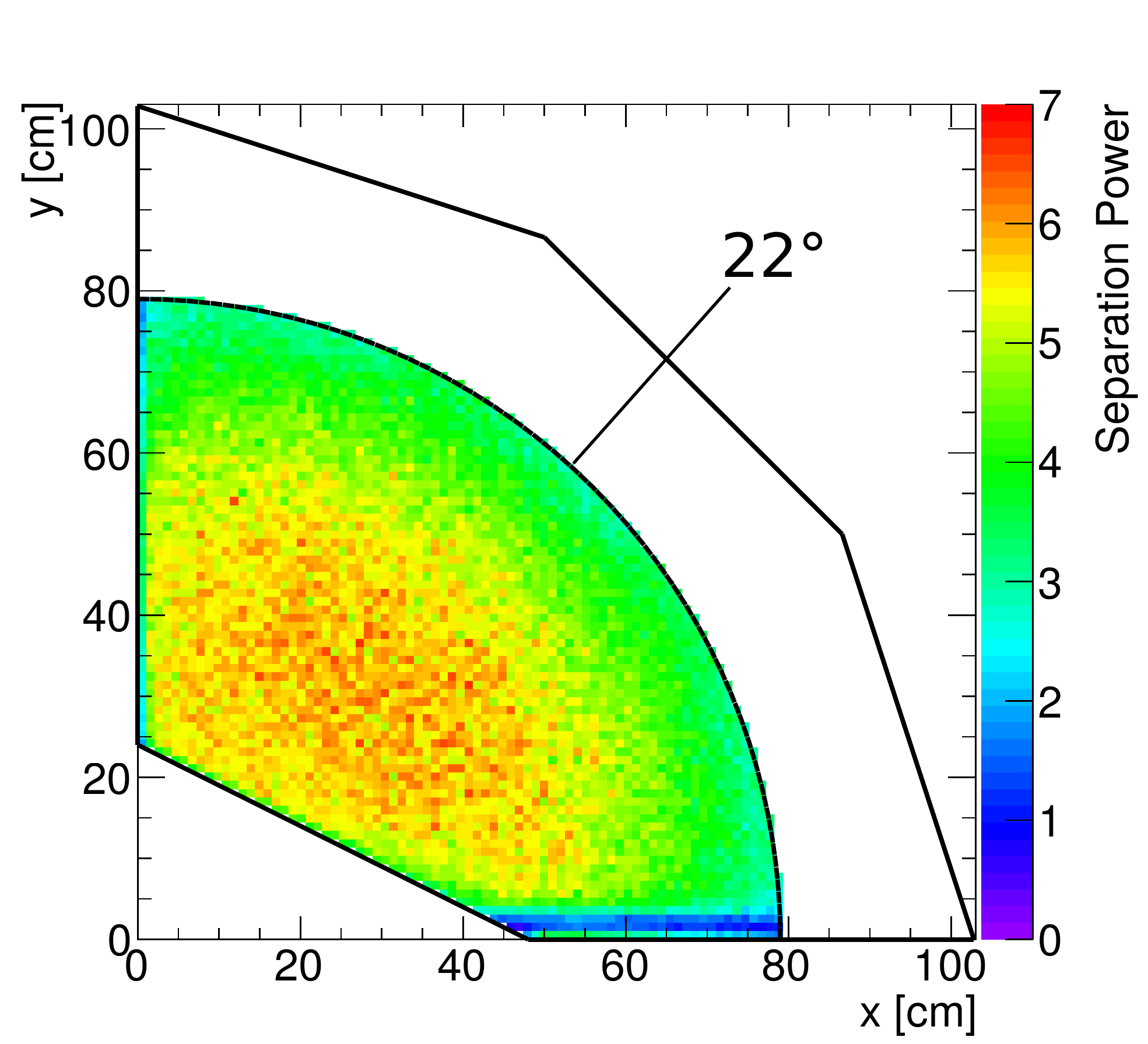}
\end{center}
\caption{\label{fig:simulation}Left: Separation power for pions/kaons at 2\,GeV/c. Right: Separation power for pions/kaons at 4\,GeV/c.}
\end{figure}
\section{Testbeam Data}
\subsection{DESY 2016}
Several testbeam campaigns have been used to validate the obtained simulation results. One successful testbeam took place in October 2016 at DESY by using a 3\,GeV/c electron beam to test the detector performance. The measured single photon resolution of angle and position scans, as well as the photon multiplicity, match very well with the MC simulations.

\begin{figure}
\begin{center}
\includegraphics[width=\textwidth]{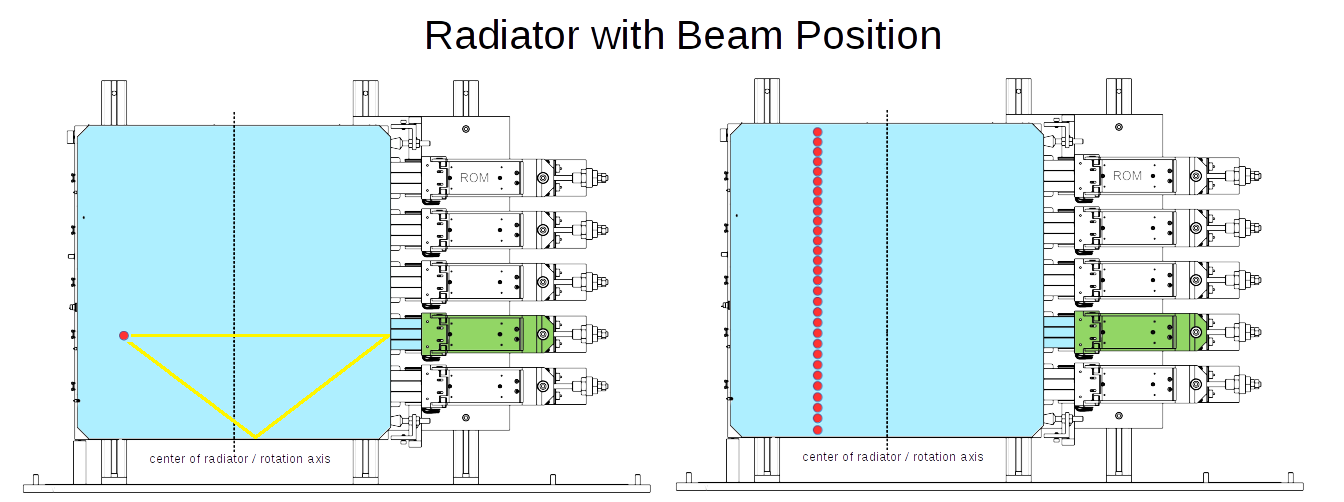}
\caption{\label{fig:testbeam}The setup for the testbeam at DESY in 2016 showing the positions of the beam and ROM for different angle scans (left) and position scans (right).}
\end{center}
\end{figure}

In contrast to the final detector design, a prototype of the radiator plate with a size of $50\times 50\,\mathrm{mm}^2$ was assembled in a newly designed and light-tight box. For the first time, the TOFPET readout system was used for data acquisition. During the beamtime, one full ROM with 3 FELs and 1 MCP-PMT was attached to the radiator plate.  A sketch of the prototype setup with the used radiator plate, the attached ROM, and the beam positions of the angle and vertical position scan can be found in Figure~\ref{fig:testbeam}.

For the execution of the vertical position scan, the hit pattern for a series of 30 different positions has been acquired. The distances between these data points was set to 17\,mm, in order to match them with the desired bar width of 16\,mm and an additional gap of 1\,mm between two bars. In the next step, each single event of the 30 beam positions was combined with the remaining 29 positions to a new event. This so called \textit{event combination} method allows the computation of the average resolution for a EDD with 30 \textit{virtual} FELs attached to the radiator plate. However, some further systematic errors of this method, like e.g. angle straggling of the charged particle, might deteriorate the final results slightly but have not been further studied.

\begin{figure}
\begin{center}
\includegraphics[width=0.49\textwidth]{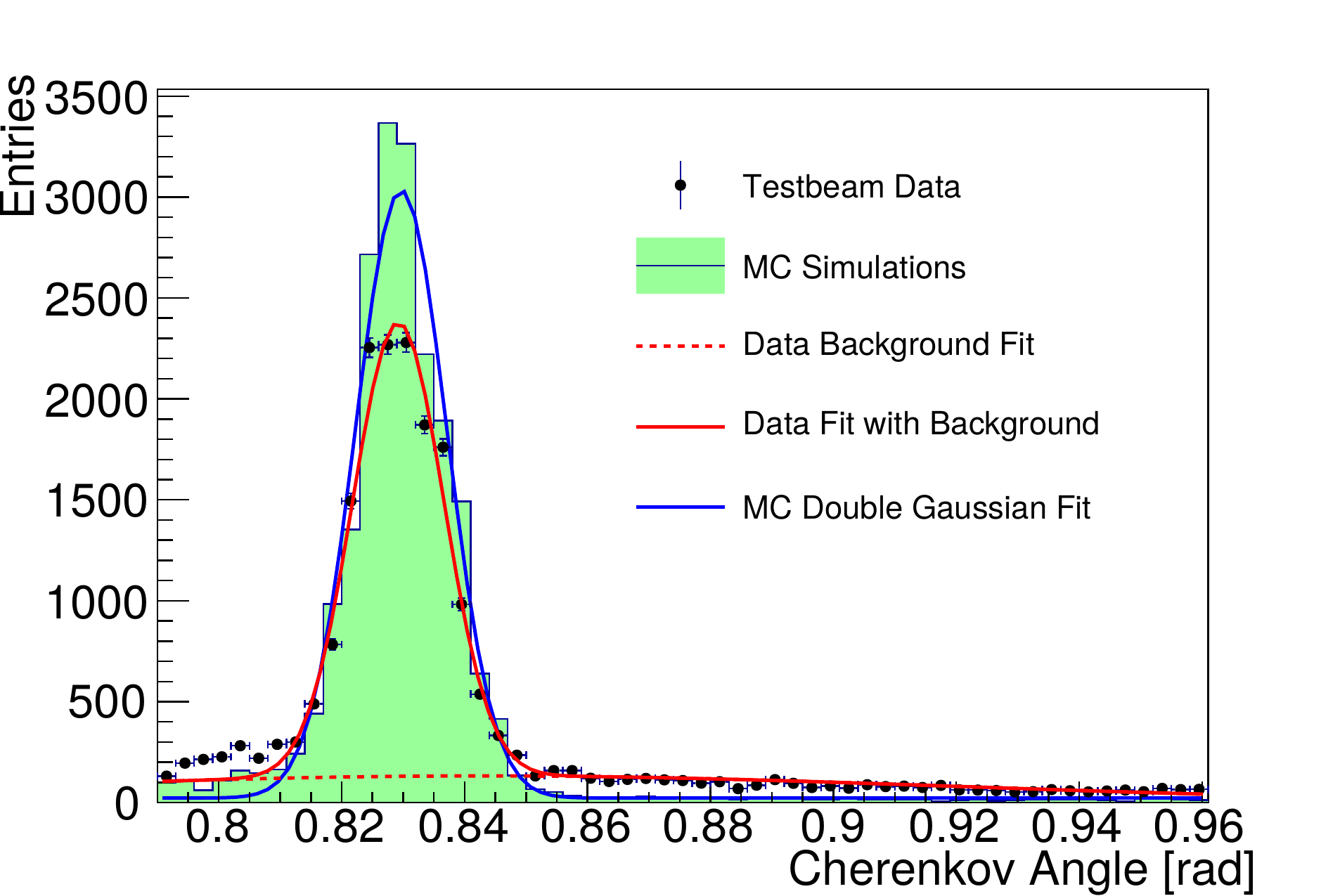}
\includegraphics[width=0.49\textwidth]{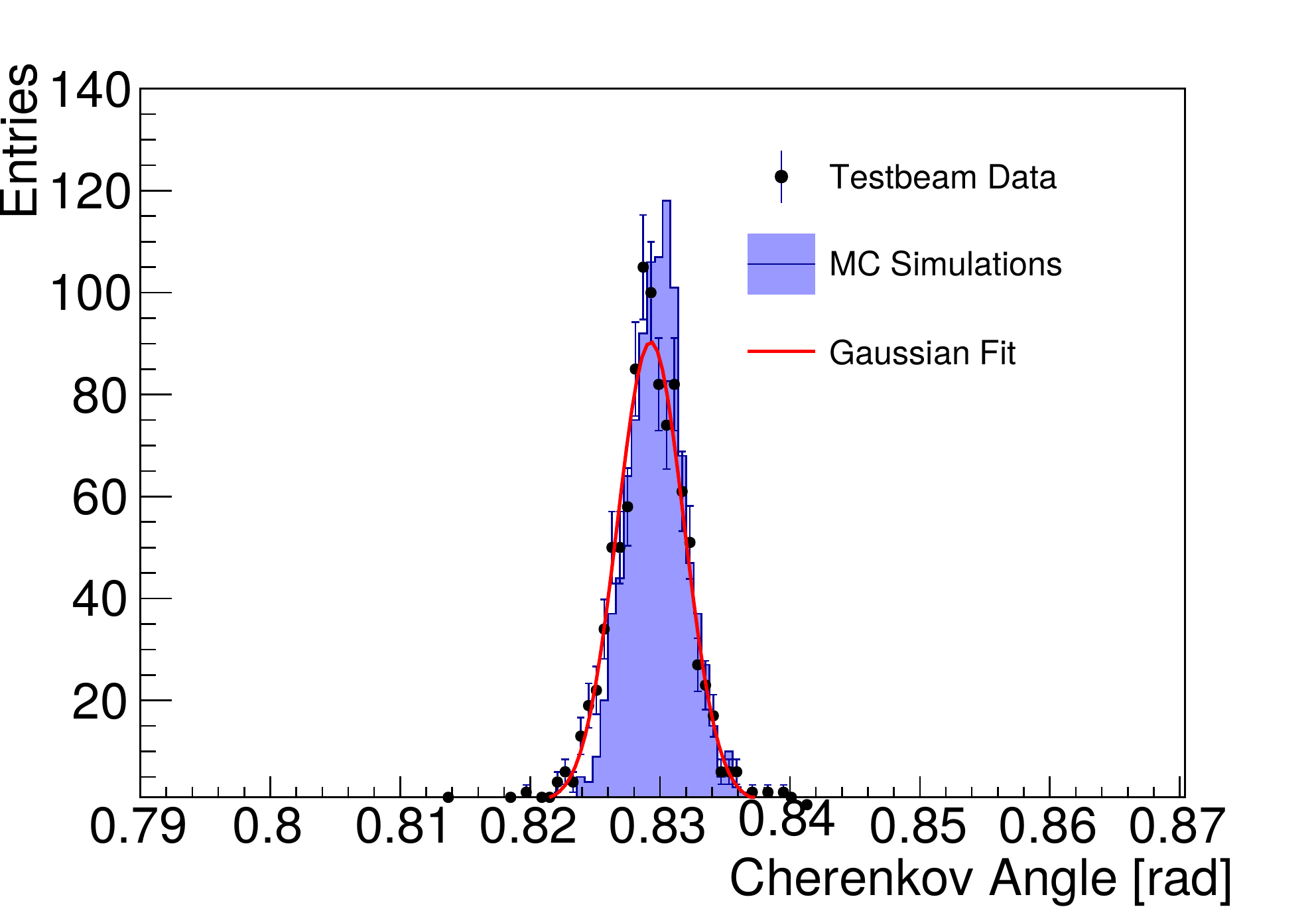}
\caption{\label{fig:combination}Left: The single photon resolution of 30 positions during a vertical position scan. Right: The detector resolution obtained from an event combination of all 30 positions.}
\end{center}
\end{figure}

The obtained single photon resolution with the center FEL of the attached ROM is shown on the left side of Figure~\ref{fig:combination} and matches very well with the MC simulations that have been performed with a standalone EDD simulation framework using the standard Geant4 libraries. For the event combination, an averaging algorithm, that is based on the truncated mean method, has been used to filter out background hits and increase the detector resolution. The right side of Figure~\ref{fig:combination} shows the combinations of the reconstruction results of the MC studies and the measured data. It turned out that the obtained resolution of 2.5\,mrad is slightly worse than the requirement of the final detector. This difference can be explained with the specifications of the installed MCP-PMT, that differ from the one in the final detector, and the absence of an optical filter. Furthermore, additional effects regarding a lower beam momentum and different particle species have to be taken into account. Despite that, the measured photon yield matches very well with the performed MC simulations.
\subsection{CERN 2018}
In order to validate the obtained results with a hadronic beam, a new testbeam campaign had been scheduled to August 2018 and ran in the T9 area of CERN with a proton/pion provided by collisions of protons with a hadron target behind the Proton Synchrotron (PS). The testbeam setup of the EDD prototype is very similar to the one at DESY. However, in contrast to the previous testbeam campaign, a new version of the TOFPET ASICs had been used and additionally, a new type of MCP-PMT photocathode with an enhanced QE in in the green area was attached to one of the 3 fully equipped ROMs.

\begin{figure}
\begin{center}
\includegraphics[width=0.9\textwidth]{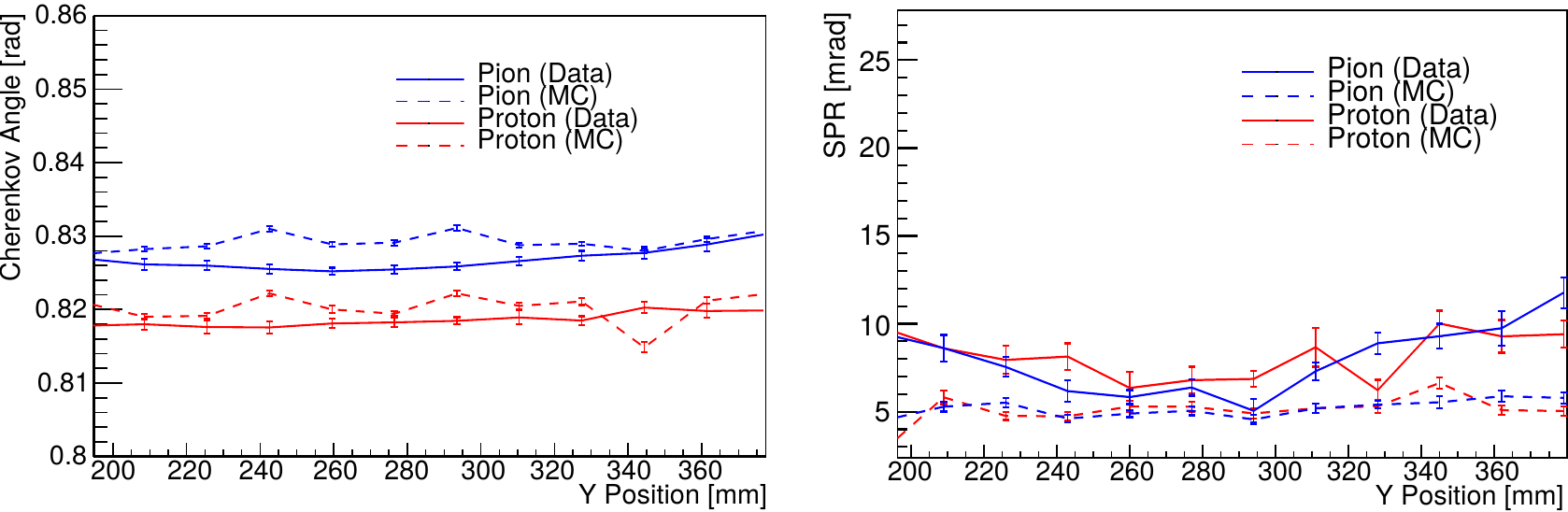}
\end{center}
\caption{\label{fig:posscan}The Cherenkov angle reconstruction (left) and single photon resolution (right) of the data taken during a testbeam at CERN in 2019.}
\end{figure}

The first preliminary results of a performed vertical scan, that is similar to the one at DESY, are shown in Figure~\ref{fig:posscan}. These results have been obtained at a beam momentum of 7\,GeV/c since this momentum corresponds to a pion/kaon separation at around 3.5\,GeV/c because of the differing particle rest masses of protons and kaons. By using the time information of two additionally installed TOF counters, that were placed in the particle beam with a distance of 29\,m to each other, an external separation of pions and protons allowed to study the Cherenkov angle and single photon resolution of different particle species separately. A deeper analysis indicated that the misidentification rate was around 2\% for both particle species. However, this method can only be applied for particle momenta below 10\,GeV/c because for higher beam momenta the TOF signals cannot be separated sufficiently.

As shown above, the reconstructed Cherenkov angle fits to the one of the MC predictions and is close to the theoretical value. The single photon resolution matches very well with the MC simulation results at a position around 300\,mm. This position is identical to the one of the analyzed FEL and therefore expected to deliver the best performance. The influences of possible misalignments in the optics or other factors, that might have deteriorated the resolution, are currently under investigation.

The slightly worse resolution at other positions can be explained with the detection of fewer photons because of geometrical effects. This increasing inefficiency for larger $y$ values results in a smaller signal-to-background ratio. An additional problem was found in the application of the new ASIC version which were officially not compatible to pulses with a negative polarity. Hence, the detector efficiency had been worse than expected, leading to the result of a much smaller photon yield compared to MC simulations. Together with an enlarged photon background and problems with several hot and dead channels due to issues with the chosen threshold settings, the signal shape was severely distorted for some of the data points and could merely be used for signal reconstructions. Nevertheless, the performance of the setup turned out to be much better than estimated during the data acquisition and will be investigated further in the near future.
\section{Conclusion \& Outlook}
The Endcap Disc DIRC is a new type of DIRC detector that has not been used for any other high energy experiment (HEP) before. It can be concluded from the previously shown simulation studies, that the EDD performance show very promising results over the full coverable kaon phase space. Various test beam campaigns have then been used to validate these results. New ASICs of the new TOFPET version, that do support pulses with a negative polarity, will be delivered during the following months by the production company. With these new ASICs, further testbeam campaigns and measurements with cosmic muons will be used to finalize the detector design.

The second phase of PANDA with a fully equipped target and forward spectrometer will start with data taking in 2027 according the actual PANDA time schedule. Thus, the production of the full EDD has to be finished around 2026. However, it is planned to install already one quadrant for testing purposes in the first phase of PANDA, that is going to start in 2024 by commissioning with a proton beam, to validate the performance under the conditions of the PANDA detector environment.
\bibliographystyle{unsrt}
\bibliography{proceedings}

\begin{thebibliography}{1}

\bibitem{DESTEFANIS2013199}
M.~Destefanis.
\newblock {The PANDA experiment at FAIR}.
\newblock {\em Nuclear Physics B - Proceedings Supplements}, 245:199 -- 206,
  2013.
\newblock The Proceedings of the 7th Joint International Hadron Structure'13
  Conference.

\bibitem{Schmidt:2018fbc}
M.~Schmidt et~al.
\newblock {Endcap Disc DIRC for PANDA at FAIR}.
\newblock {\em Springer Proc. Phys.}, 212:275--278, 2018.

\bibitem{Schwarz:2019law}
C.~Schwarz et~al.
\newblock {The Barrel DIRC detector of PANDA}.
\newblock {\em Nucl. Instrum. Meth.}, A936:586--587, 2019.

\bibitem{phd:etzelmueller}
Erik Etzelm\"uller.
\newblock {\em {Developments towards the technical design and prototype
  evaluation of the PANDA Endcap Disc DIRC}}.
\newblock PhD thesis, University of Giessen, Gie\ss{}en, 2017.

\bibitem{paper:cherenkov}
Pavel~A. Cherenkov.
\newblock {Visible emission of clean liquids by action of $\gamma$ radiation}.
\newblock {\em Doklady Akademii Nauk}, 2(451), 1934.

\bibitem{article:tofpet}
M.~D. Rolo, R.~Bugalho, F.~Goncalves, G.~Mazza, A.~Rivetti, J.~C. Silva,
  R.~Silva, and J.~Varela.
\newblock {TOFPET ASIC for PET applications}.
\newblock {\em Journal of Instrumentation}, 8(02):C02050, 2013.

\bibitem{Kuhn:2017bbd}
Wolfgang K\"uhn, S\"oren Lange, Yutie Liang, Zhen-An Liu, Simon Reiter, Milan
  Wagner, and Jingzhou Zhao.
\newblock {Data Acquisition and Online Event Selection for the PANDA
  Experiment}.
\newblock {\em JPS Conf. Proc.}, 18:011035, 2017.

\bibitem{article:geant4}
S.~Agostinelli et~al.
\newblock Geant4 -- a simulation toolkit.
\newblock {\em Nuclear Instruments and Methods in Physics Research Section A:
  Accelerators, Spectrometers, Detectors and Associated Equipment}, 506(3):250
  -- 303, 2003.

\bibitem{article:spataro_pandaroot}
Stefano Spataro and the PANDA~Collaboration.
\newblock The pandaroot framework for simulation, reconstruction and analysis.
\newblock {\em Journal of Physics: Conference Series}, 331(3):032031, 2011.

\end{thebibliography}
\end{document}